\documentclass[fleqn,10pt]{wlscirep}
\usepackage[utf8]{inputenc}
\usepackage[T1]{fontenc}
\usepackage{braket}
\usepackage{multirow}
\usepackage{footnote}
\usepackage{tablefootnote}
\usepackage{float}
\usepackage{graphicx}
\usepackage{amsmath}
\usepackage{xcolor}
\usepackage{cite}
\usepackage{multirow}
\usepackage{hyperref}
\usepackage{amsfonts}
\usepackage{bbold}
\usepackage{subcaption}
\title{Combating quantum errors: an integrated approach}

\author[1*]{Rajni Bala}
\author[1]{Sooryansh Asthana}
\author[1]{V. Ravishankar}
\affil[1]{Department of Physics, Indian Institute of Technology Delhi, New Delhi-110016, India}

\affil[*]{Rajni.Bala@physics.iitd.ac.in}

\begin{abstract}

Near-term quantum communication protocols suffer inevitably from channel noises, whose alleviation has been mostly attempted with resources such as multiparty entanglement or sophisticated experimental techniques. 
Generation of multiparty higher dimensional entanglement is not easy.  This calls for exploring realistic solutions which are implementable with current devices. Motivated particularly by the difficulty in generation of multiparty entangled states, in this paper, we have investigated error-free information transfer with minimal requirements. For this, we have proposed a new information encoding scheme for communication purposes. The encoding scheme is based on the fact that most noisy channels leave some quantities invariant. Armed with this fact, we encode information in these invariants. These invariants are functions of expectation values of operators.  This information passes through the noisy channel unchanged. Pertinently, this approach is not in conflict with other existing error correction schemes. In fact, we have shown how standard quantum error-correcting codes emerge if suitable restrictions are imposed on the choices of logical basis states. As applications, for illustration, we propose a quantum key distribution protocol and an error-immune information transfer protocol. 
\end{abstract}
\begin{document}

\flushbottom
\maketitle

\thispagestyle{empty}

\section*{Introduction}
\label{intro}
The last three decades have witnessed  a burgeoning interest in study of quantum communication, quantum computation and, quantum search, to name a few \cite{Bennett84, Bennett92, Ekert91, Deutsch92, grover1996fast}. The interest largely owes to a promise of outperforming their classical counterparts, or proposals of altogether novel applications not possible with purely classical resources. However, their implementations inevitably demand generation and transfer of quantum states, applications of various quantum gates and measurements, all within their coherence times. 

Noise unveils itself as a major impediment in meeting with these demands \cite{chubb2019noise}. In order to mitigate the effect of noise on quantum protocols, quantum error correcting codes have already been proposed in the seminal work of Shor and Steane \cite{Shor1995, steane1996simple}. In fact, various representations of the Knill-Laflamme condition \cite{knill1997theory}  have been obtained, which serve as error-correcting codes for different noisy models \cite{layden2019ancilla,BenyPRL,PhysRevA.76.042303}. These codes, in turn, require multiparty entanglement, which is a costly resource, as is reflected in the experimental challenges faced in its generation \cite{huang2011experimental, lu2007experimental, proietti2021experimental}. In fact, for orbital angular momentum of light, which is arguably at the forefront of realisation of quantum hardware for communication protocols, there is an upper limit beyond which higher dimensional multiparty entanglement cannot be generated as yet \cite{erhard2018experimental}. As a consequence, in spite of having a large number of very promising and elegant quantum information-theoretic protocols, their experimental implementations,  those too at a scalable stage,  still pose a challenge \cite{PhysRevApplied.14.064037, pirandola2015advances, diamanti2016practical}. It has been evidenced in, for example, quantum error correction assisted quantum key distribution with higher dimensional systems 
\cite{PhysRevApplied.14.064037}.  These observations have been succinctly encapsulated in the phrase `noisy-intermediate-scale-quantum computation', coined by John Preskill \cite{preskill2018quantum} for the current stage of quantum information processing. In fact, these observations have led to several protocols and error correction schemes that are realisable with current technologies and near-term quantum devices \cite{lo2006security}.  Examples include approximate error-correcting codes \cite{crepeau2005approximate}, error-avoiding codes \cite{zanardi1997error}, probabilisitic error correction \cite{PhysRevLett.82.2598} and error-suppression techniques \cite{koczor2021exponential}, quantum error mitigation techniques \cite{cai2021quantum,Otten19}. Almost all these schemes for combating errors aim at retrieving the quantum state at the end of a noisy channel. This demand either imposes restrictions on the choice of basis states or requires repeated interventions at appropriate times, making these techniques resource costly.   Recently, recovery of noise-free observables has also been studied by employing the results of slightly different experiments \cite{Otten19}.

This prompts us to ask a question: how do we find error-combating techniques without putting any restriction on the nature of states (i.e., without using multiparty entanglement)? That is to say, we look for resource-friendly error combating techniques for protocols that aim at solely information transfer. The answer to the above question is embedded in a new proposal of modifications of encoding and decoding procedures. For, unlike classical communication, in which bits are synonymous with the information they carry, their quantum counterparts, {\it viz.}, qubits carry information in sesquilinear functions of states of these qubits. This fact indicates that information can also be encoded in expectation values of different operators which are indeed sesquilinear functions of states.  Since individual outcomes of different operators are random, their concatenated strings are also random. As a result, their collective property, {\it viz.}, average is also random. This randomness in expectation values of different operators legitimises their employment as information carriers. By extension, those combinations of expectation values that remain invariant under a noisy evolution of a state emerge as  natural candidates for transferring information in an error-immune manner.

In this work, we systematically find out the quantities that remain invariant under different noisy evolutions. These invariants involve combinations of expectation values of the operators that get rescaled in a compensatory manner under a noisy evolution of a state. The invariants may be categorised into three different families-- (i) in the first family, they are simply the expectation values of operators, (ii) the second family consists of ratios of expectation values of two operators, and, (iii) the third family consists of quantities that are combinations of expectation values of operators -- not necessarily scaled by the same factor. The advantage in this approach is that with almost any state, one gets a few quantities which remain unchanged during the noisy evolution.

In this scheme, quantumness  is manifested in two ways-- (a) in non-commuting nature of observables whose expectation values are involved, (b) in non-orthogonality of states employed for quantum communication.  These features allow for (i) transfer of more information and, (ii) for security against eavesdropping in secure quantum communication protocols. As concrete applications, we propose two quantum communication protocols falling in the two categories-- (i) quantum communication without security against eavesdropping, (ii) secure quantum communication. These protocols act as templates for a large family of  other protocols, which can be proposed for error-free information transfer through noisy channels in several other scenarios.

 At this juncture, we wish to point out that this approach is not orthogonal to the conventional quantum error-correcting codes (QECC). Contrarily, we show that if the basis states are chosen to  satisfy the celebrated Knill-Laflamme condition, the conventional error-correction codes result. Since the latter have been thoroughly studied, we do not elaborate on them in this work except for showing an interrelation between QECC and the approach proposed here. What this study serves to unravel is a nice interplay of consumption of resources in the techniques of error mitigation and consequent retrieval of information or state.

The plan of the paper is as follows:  in ``\hyperlink{Notation}{Notation}'', we setup the notation to be used throughout the paper. In ``\hyperlink{Setting up the stage}{Idea advanced in this work}'', we elucidate the information encoding scheme proposed in this work with an example. In  ``\hyperlink{Formalism}{The formalism}'', the central result of the paper, i.e., the formalism to obtain invariants for various noisy channels is presented. The information encoded in these invariants can be transferred without any error.  In ``\hyperlink{generalization}{Information transfer with qu$N$its}'', invariants are identified for various noisy channels of a qu$N$it. 
As an application of the encoding scheme, in ``\hyperlink{Applications}{Application: QKD employing qubits in a depolarising channel}'',  a protocol for quantum key distribution is demonstrated. The section on  ``\hyperlink{Emergence of quantum error correcting codes}{Emergence of ancilla-free quantum error correction}'' shows how error correction codes result from this framework if logical basis states are chosen appropriately. The section titled ``\hyperlink{Conclusion}{Conclusion}'' concludes the paper with closing remarks.

\section*{Notation}\hypertarget{Notation}{}\label{preliminaries}
In this section, we setup the notation to be used henceforth in the paper. 
\begin{enumerate}
    \item The set of linear operators acting on an $N$- level system forms a vector space of dimension $N^2$. The basis operators may be chosen to be:
  \begin{align}
&    S^{(kl)}\equiv |k\rangle\langle l|+|l\rangle\langle k|,~~A^{(kl)}\equiv-i(|k\rangle\langle l|-|l\rangle\langle k|),~~k > l;~~~~~   d^{(k)}\equiv |k\rangle\langle k|,~~~k,l \in \{0,1,\cdots,N-1\}
     \label{eq:off-diagonal}
  \end{align}
   For future use, the diagonal operators, $D^{(kl)}$, are defined as:
\begin{equation}\label{eq:sum_diagonal}
    D^{(kl)}\equiv d^{(k)}-d^{(l)},~~ k\neq l, ~~k, l \in \{0, \cdots, (N-1)\}.
\end{equation}
\item The set of generalised Pauli operators acting on a single qu$N$it consists of products of powers of operators $X$ and $Z$, which are defined as:
\begin{equation}\label{eq:operator_X}
    X\equiv\sum_{k=0}^{N-1} \ket{k+1}\bra{k},\quad\quad Z\equiv \sum_{k=0}^{N-1} \omega^k \ket{k}\bra{k} ,
\end{equation}
where $\omega\equiv e^{\frac{2\pi i}{N}}$ and addition of integers is considered modulo $N$.  Both the operators $X$ and $Z$ are unitary operators satisfying $X^N,~Z^N=\mathbb{1}$.
\end{enumerate}

 \section*{Idea advanced in this work}
 \label{Setting up the stage}
 \hypertarget{Setting up the stage}{}
Before laying down the formalism, we first elucidate the idea advanced in this work. An $N$- dimensional quantum state, in general, is characterised by $(N^2-1)$ independent parameters. These parameters are expectation values of $(N^2-1)$ generators of the group $SU(N)$. Hence, one may also envisage a scenario in which the expectation values of these generators contain the information. That is to say, the expectation values of these generators are to be treated as random variables. For encoding of information in these  expectation values, we must consider a probability distribution of expectation values.

As an example, consider a two-dimensional pure state, $\rho =\frac{1}{2}(\mathbb{1} +\vec{\sigma}\cdot\hat{p})$, $|\hat{p}|=1$. This state is characterised by  the three dimensional vector $(\sin \theta \cos\phi, \sin \theta\sin\phi, \cos\theta)$. Here $\theta, \phi$ are respectively the polar and azimuthal angles of $\hat{p}$ with $\theta  \in [0, \pi]$ and $\phi \in [0, 2\pi]$. The three components of this vector are expectation values of the generators of $SU(2)$, {\it viz.}, $\sigma_x, \sigma_y, \sigma_z$.  In principle, the three dimensional vector may point along any direction on the Bloch sphere, thanks to the continuous ranges of the parameters $\theta$ and $\phi$ providing randomness in the expectation values of $\sigma_x, \sigma_y, \sigma_z$.  

Though expectation values can take infinitely many values, however, in reality, infinite precision is just a  mathematical artifice. One is bound to deal with finite precisions of detectors, which limits the information content. This motivates us to employ qu$N$its, as they belong to a larger Hilbert space.   A qu$N$it is characterised by $(N^2-1)$ independent parameters, which can be determined by measuring expectation values of the generators of the group $SU(N)$. So, higher dimensional states inevitably carry more information encoded in the expectation values of $(N^2-1)$ generators.

\subsection*{Advantage of this encoding scheme}
At this juncture, it is worthwhile to consider whether the proposed information encoding scheme affords any additional advantage as compared to the conventional schemes. Interestingly, the advantage best manifests itself when one considers noisy evolutions of states. In a noisy evolution, a quantum state is inevitably changed. However, as will be shown in the subsequent sections, there exist  some combinations of expectation values that remain unchanged. Hence, this encoding scheme provides us with a niche for transfer of information in an error-immune manner via  those invariants. Additionally, this encoding scheme does not put any constraint on the purity of states. This reduces the burden of preparation of states as mixed states can also be employed.  As is hopefully clear from the discussion, this encoding-decoding scheme does not aim at retrieving the state. It serves to determine the values of invariant quantities, that too without consuming any additional resource like multiparty entanglement and sophisticated experimental techniques (like repeated interventions at precise time intervals \cite{PhysRevApplied.5.064013}).

\section*{ The formalism}
\label{formalism}
\hypertarget{Formalism}{}
In this section, we lay down the formalism to obtain invariants for noisy communication channels.  Noisy channels are well studied through  completely-positive-trace-preserving (CPTP) maps \cite{nielsen_chuang_2010}.

Let $\rho$ be the state of the system used for communication. After passing through a noisy channel, whose action is denoted by a quantum operation $\mathcal{E}$, the state of the system changes to $\rho'$, 
\begin{equation}\label{eq:final}
    \rho\rightarrow \rho' \equiv \mathcal{E}\big(\rho\big)= \sum_k E_k\rho E_k^\dagger.
\end{equation}
 The Kraus operators $E_k$ satisfy the condition, $\sum_k E_k^\dagger E_k=\mathbb{1}$. The range of $k$ specifies the number of Kraus operators required to represent a quantum channel of interest.
\subsection*{Identification of invariants}
 Under a noisy evolution of a state, the final state is, in general, mixed and involves parameters that determine the initial state as well as the noisy channel (see equation (\ref{eq:final})). Of particular significance for us is  that we can construct those quantities (mostly in the form of ratios of expectation values of observables) that are solely functions of parameters of the initial state and are independent of those of noisy evolution operators. Since, by construction,  these quantities  involve  parameters of initial state only, they remain invariant under a noisy evolution. As such they can be legitimately employed for information transfer in an error--immune manner.
 
For this, we consider those operators whose expectation values are re-scaled under a noisy evolution.  Let $\{O_\alpha\}$ be the set of such operators satisfying,
\begin{equation}
    \langle O_\alpha\rangle_{\rho'} =\lambda_\alpha \langle O_\alpha\rangle_\rho,
    \label{Evolution}
\end{equation}
where the non-vanishing scaling factors $\lambda_\alpha $ are functions of noise parameters. Together with equation (\ref{eq:final}), we get the following equation,
\begin{align}
    & \sum_k {\rm Tr}\big(E_k^\dagger O_\alpha E_k \rho\big)= \lambda_{\alpha}{\rm Tr}(O_{\alpha}\rho).
    \label{Evolution_op1}
\end{align}
Since equation (\ref{Evolution_op1}) holds for an arbitrary state $\rho$, the operator $O_\alpha$ must satisfy the following condition,
\begin{equation}\label{eq:op}
    \sum_k E_k^\dagger O_\alpha E_k=\lambda_\alpha O_\alpha.
\end{equation}
Suppose that there are $n$ such operators $\{O_1, \cdots, O_n\}$, satisfying equation (\ref{Evolution}). Consider the following function of expectation values,
\begin{equation}\label{eq:product}
I (\{r_{\alpha}\})\equiv \prod_{\alpha=1}^n\bigg( \langle O_\alpha\rangle_{\rho'}\bigg)^{r_\alpha}=\prod_{\alpha=1}^n \lambda_\alpha^{r_\alpha}\bigg( \langle O_\alpha\rangle_{\rho}\bigg)^{r_\alpha},
\end{equation}
where $r_\alpha$ can take real values.
The quantity,
$I(\{r_{\alpha}\})$, would remain invariant under the noisy evolution of a state, iff,
\begin{equation}\label{eq:condition}
   \prod_{\alpha=1}^n \lambda_\alpha^{r_\alpha}=1. 
\end{equation}
 In that case, the information encoded in the quantity, $I(\{r_{\alpha}\})$, remains invariant. We now categorise these invariants into three different families depending upon how the operators are re-scaled under a noisy evolution of a state. 

\subsubsection*{The first family of invariants }
The first family of invariants corresponds to the situation when,  $r_{\alpha} =1$, for a given value of $\alpha$ and all other $r_{\alpha}$'s vanish in equation (\ref{eq:product}). For this case, the condition (\ref{eq:op}) reduces to: 
\begin{equation}\label{eq:operator}
    \sum_k E_k^\dagger O_{\alpha} E_k= O_{\alpha}.
\end{equation}
 This condition implies that $\langle O_{\alpha}\rangle$ remains unchanged under a noisy evolution of a state, and hence, can be used for encoding information. Note that the equation (\ref{eq:operator}) is satisfied whenever  $[O_{\alpha}, E_k]=0,~ \forall ~k$. The range of $\alpha$ determines the number of independent operators whose expectation values remain invariant. For the case in which all the  Kraus operators are generated by a single unitary operator $U$, i.e., $E_k=U^k$, the statistics of eigenprojections of $U$ remains unchanged under the evolution.  Hence, they act as invariants.

  This particular set of invariant quantities arises in the study of Lindblad master equations and is used to recognize steady-state structures in a noisy state \cite{albert2014symmetries}.

 \subsubsection*{The second family of invariants} \label{second family}
Having discussed the invariants from the first family, it is worthwhile to ask whether there exists any other independent invariants, not exhausted in the first family. It is because there may be instances in which no invariant can be obtained from the first family. Therefore, we consider a more general case and define the second family of invariants.  This family involves ratios of expectation values of those operators, that scale identically under a noisy evolution.

To elaborate,  the second family corresponds to the  condition when only two $\lambda_{\alpha}'s$, say, $\lambda_1$ and $\lambda_2$ survive and are equal to each other. Then, the expectation values of the two operators follow the relation, $
   \langle O_1\rangle_{\rho'}=\lambda \langle O_1\rangle_{\rho},\langle O_2\rangle_{\rho'}=\lambda \langle O_2\rangle_{\rho}$. For these operators, equation (\ref{eq:product}) provides invariant when $r_1=1$ and $r_2 = -1$ and all the other $r_{\alpha}$'s vanish. This implies that the information encoded in the quantity,
  \begin{align}
  \label{eq:invariant_condition}
        I\equiv\frac{\langle O_1\rangle}{\langle O_2\rangle},
  \end{align}
 remains free from errors and can be transferred reliably. 
\subsubsection*{The third family of invariants}\label{third family}
Most of the invariants are exhausted by the first two families. However, there may exist a set of operators which satisfies the condition (\ref{eq:condition}), but does not belong to either of the aforementioned families. They constitute the third family of invariants. This family consists of those functions for which the condition (\ref{eq:condition}) is satisfied either for $n>2$ or for $n=2$ when $r_1\neq -r_2=1$ .

 \section*{Results}
 In this section, employing the formalism, we obtain invariants for various noisy channels of a qu$N$it.  This is followed by proposal of a quantum key distribution protocol employing these invariants. Finally, we also show how this formalism embeds quantum error correcting codes, contingent on suitable restrictions on choices of logical basis states. 
 \section*{Information transfer with qu$N$its}
\label{generalization}
\hypertarget{generalization}{}
Recently, quantum communication protocols employing qu$N$its have attracted a lot of interest, thanks to the experimental advances in generation, manipulation and detection of photonic qu$N$its (see, for example, \cite{shen2019optical} and references therein). The effect of various noisy channels such as generalised Pauli channels, depolarising channel, dephasing channel and amplitude damping channel  on these protocols have been studied \cite{fonseca2019high,xu2022enhancing, fortes2015fighting,fortes2016probabilistic}. Appreciating the significance of these channels, we identify invariants for the same employing the formalism laid down in the previous section. To start with, we consider the generalised Pauli channel.

\subsection*{Generalised Pauli channel}
The effect of the generalised Pauli channel (GPC) on a state can be modelled with the help of operators \cite{miller2018propagation} $X\equiv\sum_{k=0}^{N-1} \ket{k+1}\bra{k}$ and $Z\equiv \sum_{k=0}^{N-1} \omega^k \ket{k}\bra{k} .$ 
An arbitrary state $\rho$, after passing through this channel changes to the state $\rho'$,
\begin{equation}\label{eq:GP}
    \rho\rightarrow\rho'\equiv\sum_{r,s=0}^{N-1} ~p_{rs} (X^rZ^s)~\rho ~(X^rZ^s)^\dagger,~~~~~~~\sum_{r,s=0}^{N-1} p_{rs}=1,
\end{equation}
where $0\leq p_{rs}\leq 1$ is the probability with which the unitary operator $X^rZ^s$ corrupts the state. 
It is straightforward to verify that the GPC does not lead to any invariant. Since the set of Kraus operator $\{X^rZ^s\}$ forms an irreducible representation of Weyl algebra, following Schur's lemma, there cannot be any invariant subspace except the trivial identity. Hence, there cannot exist any invariant from the first family. There does not exist any invariant from the second or the third family, given the fact that  all the probabilities $p_{rs}$ are distinct. The proof for the same is given below.  
Hence GPC admits no invariant except the trivial identity.

However, there exist many special cases of GPC whose effect have been studied on various communication protocols \cite{fonseca2019high,iqbal2021analysis,hu2021novel}. We now consider these special cases of GPC.\\

\noindent {\bf Proof of nonexistence of invariants belonging to the second family in a generalised Pauli channel}\\
Consider an operator $O \equiv \sum_{m,n}c_{mn}X^mZ^n$, which evolves in a generalised channel in the following manner,
\begin{align}
O \to O'&=    \sum_{r,s} p_{rs}(X^rZ^s)^{\dagger}O(X^rZ^s).
\end{align}
Employing the relation  $ZX=\omega XZ$, the above equation assumes the following form,
\begin{align}
O'&=\sum_{r,s}p_{rs}\sum_{m,n}\omega^{nr-sm}c_{mn} X^mZ^n.
\end{align}
The above equation indicates that for arbitrary values of $p_{rs}$, there cannot exist two operators that either scale by the same factor say, $\lambda$ or scale such that $\lambda_2=\lambda_1^r$, where $\lambda_{1,2}$ represents the scaling with which the expectation values of the two operators is changed.
That is why there does not exist any invariant from the second or the third family for a generalised Pauli channel.
\subsubsection*{Generalised flip error}
\label{quNitflip}
In a generalised flip error, all the relevant Kraus operators are generated by the operator $X$. 
 Let $\rho$ be the state used to send information through such a channel. After passing through this channel, it changes to the state $\rho'$:
 \begin{equation}\label{eq:quditflip}
     \rho\rightarrow\rho'\equiv \sum_{r=0}^{N-1} p_r X^r\rho~ (X^r)^\dagger,~~~0\leq p_r\leq 1,~~~\sum_{r=0}^{N-1}p_r =1.
 \end{equation}
  Employing cyclicity property of the trace one may immediately see that the expectation values of operators, $I_{1}^{(m)}\equiv\langle X^m\rangle,$  are invariant, and thus the encoded information remains error--free. In addition, employing the relation, $ZX=\omega XZ$, we obtain the following set of invariants belonging to the second family,
 \begin{equation}
I_{2}^{(ml)}\equiv\frac{\langle Z^m\rangle}{\langle Z^mX^l\rangle},
 \end{equation}
information encoded in which remains immune to errors. Enumeration of these quantities is given below.

 Since the operators $X,Z$ and $Y=XZ$ are unitarily equivalent to each other, invariants for generalised phase (replacing $X$ in the equation (\ref{eq:quditflip}) by $Z$) and for generalised combined flip and phase error (replacing $X$ in the equation (\ref{eq:quditflip}) by $Y=XZ$) can be identified in the same manner. The invariants for these two channels are given in the table (\ref{tab:quNit}).\\

 \section*{Enumeration of invariants in a generalised flip channel}
 The operators appearing in the two sets of invariants are unitary and hence, each quantity gives two real invariants.
Therefore, to ensure the independence in the information provided by invariants, the inverse operators should be excluded. Thus, the ranges of $m$ and $l$ are to be decided accordingly. We do this for odd and even $N$ separately.
\subsubsection*{\it Odd $N$}
{Range for the first family of invariants:} $m\in\{1,2,\cdots,\frac{N-1}{2}\}$.\\
{Range for the second family:} $m\in\{1,2,\cdots,\frac{N-1}{2}\}$, and $l\in\{1,2,\cdots,N-1\}$.

\subsubsection*{\it Even $N$}
{Range for the first family:} $m\in\{1,2,\cdots,\frac{N}{2}\}$.\\
{Range for the second family:} For $m\in\{1,2,\cdots,\frac{N}{2}-1\}$, $l\in\{1,2,\cdots,N-1\}$ and whenever $m=\frac{N}{2}$, $l\in\{1,2,\cdots,\frac{N}{2}\}$.\\
The difference in the range of $m$ and $l$ for even and odd $N$ arises due to the fact that for even $N$, there is an operator, $U^{\frac{N}{2}}$, which is inverse of itself and therefore, has only one independent invariant.

Given the range of $m$ and $l$ for both even and odd $N$, it is clear that the set $\mathcal{I}_{1}^{(m)}$ provides $(N-1)$ independent quantities whereas the set $\mathcal{I}_{2}^{(ml)}$ provides $(N-1)^2$ independent quantities which is quadratic in $N$. In all, there are $(N^2-N)$ independent invariants. 
 \subsubsection*{Depolarising channel}
\label{depolar_qudit}
The effect of a depolarizing channel on a state is to incoherently mix the state with white noise with a nonzero probability. The effect of this channel   on various quantum communication protocols, for example, (i) on quantum teleportation \cite{fonseca2019high}, (ii) for a qutrit \cite{xu2022enhancing}, (iii) on the key rate and hence, on security of quantum key distribution protocol \cite{iqbal2021analysis}, and,   (iv) on quantum secret sharing protocol \cite{hu2021novel} have been studied.

Let $p$ be the probability with which white noise is incoherently mixed with a state.  Let $\rho$ be the state of a qu$N$it used to transmit information through a depolarizing channel. After passing through the channel, the state of the system changes to \cite{wilde_2017}: \begin{equation}
\rho\rightarrow\rho'\equiv(1-p)\rho +p\dfrac{\mathbb{1}}{N},~~~~~0\leq p\leq 1.
\end{equation} 
The invariants, in which information can be encoded for error-free transmission are, \begin{equation}\label{eq:depolar_quNit1}
I_{1}^{(kl)}\equiv\dfrac{\langle S^{(kl)}\rangle}{\langle D^{(kl)}\rangle},~~~~~~~~~~I_{2}^{(kl)}\equiv\dfrac{\langle A^{(kl)}\rangle}{\langle D^{(kl)}\rangle},~~~~~~~~k>l;~~~~~~~~I_3^{(m)}\equiv \frac{\langle d^{(m)}\rangle-\frac{1}{N}}{\langle d^{(0)}\rangle-\frac{1}{N}},~~~~~~~~~  1\leq m \leq (N-2).
\end{equation}  

 \subsection*{Dephasing channel}
 \label{dephasing_channel}
In a dephasing channel, coherence of a state decreases without any change in its population.
Let $\rho$ be the state of a quNit system used to transfer information. After passing through a dephasing channel, it changes to \cite{marques2015experimental}:
\begin{equation}
    \rho\rightarrow\rho'\equiv\sum_{j=0}^N p_j ~E_j\rho E_j^\dagger,~~~~~~~~ 0\leq p_j \leq 1, ~~~~\sum_{j=0}^{N}p_j=1,
\end{equation}
where the relevant Kraus operators are $E_j =\mathbb{1}-2\ket{j}\bra{j}~$  ($0\leq j \leq N-1$) and $E_N=\mathbb{1}$. Under this evolution, the entries of the density matrix change in the following manner,
\begin{align}
    \rho'_{ii} =\rho_{ii}, ~~\forall i,~~~~~~~
    \rho'_{ij} =(1-2p_i-2p_j)~ \rho_{ij}, ~~~~~~\forall i \neq j. 
    \label{Relation}
\end{align}
Employing the relations given in (\ref{Relation}),  the invariants in which information can be encoded for error-free transmission are:
\begin{equation}
    \label{eq:inv_quNit_dephase}
    I^{(k)}_{1}\equiv\langle D^{(k+1,k)}\rangle,~~~~~~I_{2}^{(kl)}\equiv\frac{\langle S^{(kl)}\rangle}{\langle A^{(kl)}\rangle},~~~~~~k>l.
\end{equation}

\subsection*{Amplitude damping channel (ADC)}
In this section, we obtain a set of invariants for a qu$N$it passing through an amplitude damping channel.
 The effect of an amplitude-damping channel  on several communication protocols such as quantum teleportation, secret sharing protocol, entanglement swapping protocol and quantum key distribution protocols has  been studied \cite{fonseca2019high,hu2021novel,hu2022conclusive,iqbal2021analysis,im2021optimal,travnivcek2020experimental}.

An arbitrary state $\rho$, after passing through an amplitude damping channel, changes to the state $\rho'$: 
\begin{equation}
    \rho\rightarrow\rho'=\mathcal{E}(\rho)\equiv E_0\rho E_0^\dagger +\sum_{m<n=0}^{N-1} E_{mn}\rho E_{mn}^\dagger.
\end{equation}
The relevant Kraus operators are given as \cite{chessa2021quantum}: 
\begin{align}
    &E_0\equiv \ket{0}\bra{0}+\sum_{n=1}^{N-1} \sqrt{1-\xi_n}\ket{n}\bra{n},~~~~~~~E_{mn}\equiv\sqrt{\gamma_{nm}}\ket{m}\bra{n},
\end{align}
where $\gamma_{nm}$ describes the rate with which population from the $n^{{\rm th}}$ level is transferred to the $m^{{\rm th}}$ level. The conditions of complete positivity and trace preserving nature of the channel translates to the following inequalities: $0 \leq \gamma_{nm} \leq 1,~~ \xi_n \equiv \sum_{0 \leq m <n }\gamma_{nm}\leq 1,~\forall~ m, n~~ {\rm s.t.}~ 0 \leq m < n\leq N-1.$

The set of invariants for this channel is given by the second  and the third family as follows:
\begin{equation}
    I^{(kl)}_{1}= \frac{\langle S^{(kl)}\rangle}{\langle A^{(kl)}\rangle}, ~ ~~k>l;~~~~ ~~~~~~
    I_{2}= 
    \frac{\langle S^{(N-1,0)}\rangle\langle A^{(N-1,0)}\rangle}{\langle\pi_{N-1}\rangle},~~~{\rm where}~\pi_{N-1}\equiv \ket{N-1}\bra{N-1}.
\end{equation}

We, now, summarise the results obtained for various noisy channels of a qu$N$it succinctly in table (\ref{tab:quNit}). The salient features of the table are given below:
\begin{table}[htb!]
\begin{center}
    \begin{tabular}{|c|c|c|c|c||c|}
    \hline
    \multirow{3}{2cm}{{\bf Noisy channel}}&\multirow{2}{1.5cm}{{\bf First family of invariants }}&\multirow{2}{1.5cm}{{\bf Number of Invariants}} & \multirow{3}{2cm}{{\bf Second and third families of invariants}}&\multirow{3}{1.5cm}{{\bf Number of invariants}}&\multirow{3}{1.5cm}{{\bf Total number of invariants}} \\
 & &&&&\\
 &  && &&
 \\
 &&&&&\\
 \hline\hline
        \multirow{3}{2cm}{Generalised flip error}& \multirow{3}{*}{$\langle X^m\rangle$} &\multirow{3}{*}{$N-1$}& \multirow{3}{*}{$\dfrac{\langle Z^m\rangle}{\langle Z^m X^l\rangle}$}& \multirow{3}{*}{$(N-1)^2$}&\multirow{3}{*}{$N(N-1)$}\\
        &&&&&\\
        &&&&&\\\hline
         \multirow{3}{2cm}{Generalised phase-error }& \multirow{3}{*}{$\langle Z^m\rangle$}&\multirow{3}{*}{$N-1$} & \multirow{3}{*}{$\dfrac{\langle X^m\rangle}{\langle X^mZ^l\rangle}$}&\multirow{3}{*}{$(N-1)^2$}&\multirow{3}{*}{$N(N-1)$}\\
         &&&&&\\
         &&&&&\\\hline
          \multirow{3}{2cm}{Generalised combined flip and phase errors}& \multirow{4}{*}{$\langle Y^m\rangle$}&\multirow{4}{*}{$N-1$} &\multirow{4}{*}{$\dfrac{\langle Z^m\rangle}{\langle Z^mY^l\rangle}$}&\multirow{4}{*}{$(N-1)^2$}&\multirow{4}{*}{$N(N-1)$}\\
          &&&&&\\
          &&&&&\\
        &&&&&\\\hline
                \multirow{3}{2cm}{Dephasing channel}& \multirow{3}{*}{$\langle D^{(k,k+1)}\rangle$}&\multirow{3}{*}{$N-1$}&\multirow{3}{*}{$\dfrac{\langle S^{(kl)}\rangle}{\langle A^{(kl)}\rangle}$}&\multirow{3}{*}{$\binom{N}{2}$}&\multirow{3}{*}{$\frac{(N-1)(N+2)}{2}$}\\
        &&&&&\\
        &&&&&\\\hline
         \multirow{3}{2cm}{Depolarizing channel}& \multirow{3}{*}{--}& \multirow{3}{*}{--}& \multirow{3}{*}{$\dfrac{\langle S^{(kl)}\rangle}{\langle D^{(kl)}\rangle},~\dfrac{\langle A^{(kl)}\rangle}{\langle D^{(kl)}\rangle},\frac{\big\langle d^{(m)}\big\rangle - \frac{1}{N}}{\big\langle d^{(0)}\big\rangle-\frac{1}{N}}$}&\multirow{3}{*}{$\binom{N}{2},\binom{N}{2},N-2$}&\multirow{3}{*}{$N^2-2$}\\
        &&&&&\\
        &&&&&\\\hline
           \multirow{3}{2cm}{ADC }& \multirow{3}{*}{--}&\multirow{3}{*}{--}&\multirow{3}{*}{$\dfrac{\langle S^{(kl)}\rangle}{\langle A^{(kl)}\rangle}$, $ \dfrac{\langle S^{(N-1,0)}\rangle\langle A^{(N-1,0)}\rangle}{\langle \pi_{N-1} \rangle}$}&\multirow{3}{*}{ $\binom{N}{2},1$}&\multirow{3}{*}{ $\binom{N}{2}+1$}\\
        &&&&&\\
        &&&&&\\
        \hline
         \end{tabular}
\end{center}
\caption{ The sets of invariants for various noisy channels of a qu$N$it.}
    \label{tab:quNit}
\end{table}
\begin{enumerate}
  \item The invariants belonging to the first family are limited in number in comparison to the second family. 
  This owes to the fact that there are a larger number of expectation values that change in a scaled manner in comparison to the ones that do not change at all.
    \item  The third family provides an invariant for an amplitude--damping channel. It is  because amplitude damping channel is the sole channel  (among the ones studied in this paper) in which entries of the density matrix change with different powers of noise parameters.
    \item The increase in the number of invariants with $N$ (i.e., the dimension of the qu$N$it) further strengthens the observation that larger information can be transferred with higher--dimensional states.
\end{enumerate}

Table (\ref{tab:quNit}) also gives insight into the relative impact of noisy channels.  
In the noisy channels such as the generalised flip error, phase error, and combination of flip and phase error, the loss of information is $O(N)$. Whereas, in the channels such as dephasing and ADC, the loss of information is $O\big(\frac{N^2}{2}\big)$. 
For the sake of better elucidation,  we recall that information is encoded in the expectation values of operators or functions thereof. In a noiseless case, there are $(N^2-1)$ independent expectation values that characterize a state and hence carry information. However, for transferring error-free information, it is encoded in the invariants whose number is less than $(N^2-1)$. This decrement in the number of invariants is accounted as loss of information.

Appreciating the employment of qubits in communication protocols, the invariants for qubits passing through different noisy channels have been explicitly given below.
This concludes our discussion of invariants in various noisy channels. Following a similar procedure, the invariants for any channel can be straightforwardly obtained.

 \section*{Invariants for a qubit}
\label{Invariants_qubits}

 In this section, we have considered various noisy channels for a qubit, which even as of today, are predominantly used for communication protocols such as quantum key distribution, quantum teleportation, quantum secret sharing. In particular, we have considered a bit-flip channel, a phase-flip channel, a depolarization channel, and an amplitude--damping  channel,  and have identified invariants for the same. These noisy channels are significant not merely from theoretical considerations, but also from an experimental viewpoint. The effect of these noisy channels on various communication protocols such as quantum teleportation, superdense coding, quantum state sharing has been studied\cite{fortes2015fighting,fortes2016probabilistic,shadman2010optimal,li2019enhanced,wang2015secret}. 

  We first recall the set of Kraus operators for each of the noisy channel considered here.\\ 
\begin{itemize}
    \item Bit flip channel:  $E_1=\sqrt{1-p}~\mathbb{1}, ~E_2=\sqrt{p}~\sigma_x$.
\item Phase flip channel: $E_1=\sqrt{1-p}~\mathbb{1}$,  $E_2=\sqrt{p}~\sigma_z$.
\item Combined bit and phase flip: $E_1=\sqrt{1-p}~\mathbb{1},~ E_2=\sqrt{p}~\sigma_y$.
\item Depolarising channel: $E_1=\sqrt{1-\frac{3p}{4}}\mathbb{1},~E_2=\sqrt{\frac{p}{4}}\sigma_x,~E_3=\sqrt{\frac{p}{4}}\sigma_y,~E_4=\sqrt{\frac{p}{4}}\sigma_z$.
\item Amplitude damping channel:
\begin{align}
E_0=\dfrac{1}{2}\Big((1+\sqrt{1-q})\mathbb{1}+(1-\sqrt{1-q})\sigma_z\Big),\quad
E_1=\sqrt{q}\frac{\sigma_+}{2};\quad q\in [0,1].\nonumber    
\end{align}

\item Generalised amplitude damping channel: 
\begin{align}
 & K_1=\dfrac{\sqrt{p_1}}{2}\Big((1+\sqrt{1-q})\mathbb{1}+(1-\sqrt{1-q})\sigma_z\Big),\quad
K_2=\frac{\sqrt{p_1q}}{2}\sigma_+;\nonumber\\
& K_3=\dfrac{\sqrt{p_2}}{2}\Big((1+\sqrt{1-q})\mathbb{1}-(1-\sqrt{1-q})\sigma_z\Big),\quad
K_4=\dfrac{\sqrt{p_2q}}{2}\sigma_-,~
{\rm where}~ q\in [0,1],~ {\rm and} ~p_1+p_2=1.\nonumber
\end{align}
\end{itemize}
The invariants for all these noisy channels are identified in the table (\ref{tab:qubit}). 

\begin{table}[htb!]
\begin{center}
\begin{tabular}{|c| c |c|}
\hline
\multirow{2}{*}{{\centering\bf Noisy channel}}&\multirow{2}{2cm}{\centering\bf First family of invariants} &\multirow{2}{2.5cm}{{\bf Second and third families of invariants}}\\
 & &\\
\hline\hline
\multirow{2}{2.5cm}{ Bit-flip channel }
&\multirow{2}{*} {$\langle\sigma_x\rangle$}& \multirow{2}{*}{$\dfrac{\langle \sigma_y\rangle}{\langle\sigma_z\rangle}$} \\
&&\\
\hline
  \multirow{2}{2.5cm}{Phase-flip channel}& \multirow{2}{*}{$\langle\sigma_z\rangle$} & \multirow{2}{*}{$\dfrac{\langle \sigma_x\rangle}{\langle\sigma_y\rangle}$}\\ 
  & &\\\hline
\multirow{2}{3.5cm} {Combination of bit and phase-flip channel} & \multirow{2}{*}{$\langle\sigma_y\rangle$} & \multirow{2}{*}{$\dfrac{\langle \sigma_x\rangle}{\langle\sigma_z\rangle}$}\\
& &\\\hline 
 \multirow{2}{*}{ Depolarizing channel}
& \multirow{2}{*}{--} &  \multirow{2}{2cm}{~~~~$\dfrac{\langle \sigma_x\rangle}{\langle\sigma_z\rangle}, \dfrac{\langle \sigma_y\rangle}{\langle\sigma_z\rangle}$}\\
&&\\\hline
 \multirow{2}{2.5cm}{ Amplitude damping channel}
 & \multirow{2}{*}{--}& \multirow{2}{*}{$\dfrac{\langle \sigma_x\rangle}{\langle\sigma_y \rangle}$,~$\dfrac{\langle\sigma_x\rangle\langle \sigma_y\rangle}{\langle\pi_z^-\rangle}$}\\
&&\\\hline
\multirow{3}{3cm}{ Generalised amplitude damping channel}
  & \multirow{3}{*}{--}& \multirow{3}{*}{$\dfrac{\langle \sigma_x\rangle}{\langle\sigma_z\rangle}$}\\
&&\\ 
&&\\
\hline
 \end{tabular}
\caption{ Sets of invariants for various noisy channels of a qubit.}
    \label{tab:qubit}
    \end{center}
\end{table}
This concludes our discussion of invariants for a qubit passing through different noisy channels. 

\section*{Application: QKD employing qubits in a depolarising channel}
\label{applications}
\hypertarget{Applications}{}
Having laid down the framework, in this section, we demonstrate a protocol for quantum key distribution employing qubits passing through depolarising channel. The invariants explicitly found for a qubit passing through a depolarising channel are employed for transferring information. 
The generalisation of the protocol to quNits and to various noisy channels is straightforward.
The only difference would be in the set of invariants that are employed, as depicted in table (\ref{tab:quNit}).

  In a depolarising channel for qubits, a state $\rho$ evolves as\cite{nielsen_chuang_2010},
 \begin{align}
\rho\to (1-p)\rho+p\frac{\mathbb{1}}{2}.     
 \end{align}
There are two invariants for this channel, given by,
\begin{equation}\label{eq:QSDC_qubit1}
    I_1=\frac{\langle\sigma_x \rangle}{\langle\sigma_z \rangle},~~~~~~I_2=\frac{\langle\sigma_y \rangle}{\langle\sigma_z \rangle}.
\end{equation}
 The security of the protocol is assured by employing decoy states. The decoy states are four in number and may be chosen to be,
$    \frac{1}{2}(\mathbb{1}\pm\vec{\sigma}\cdot\hat{p}),~~\frac{1}{2}(\mathbb{1}\pm\vec{\sigma}\cdot\hat{p}'),$
where $\hat{p}$ and $\hat{p}'$ are fixed. 
With this proviso, the steps for the protocol are listed as follows:
\begin{enumerate}
    \item Alice prepares $N$ copies of each of $k$ different states, $\rho_1, \cdots, \rho_k$, to transfer information to Bob. The value of $k$ depends upon the length of the information to be transmitted. The value of $N$ is so chosen that there is sufficient statistics to determine expectation values. 
    She sends these states to Bob in a random sequence. Intermittently, she inserts a decoy state in that sequence and sends it to Bob.
    \item On receiving a state, Bob measures randomly one of the observables $\sigma_x,\sigma_y,$ and $\sigma_z$. This constitutes a round. This process is repeated for many rounds.
    \item After a sufficient number of rounds, Alice reveals the rounds in which she had sent the decoy states.  Bob reveals the observables and the corresponding outcomes for these rounds.  Alice uses this information to check for the presence of an eavesdropper, if any.
    \item If Alice is convinced about the absence of an eavesdropper, she reveals to Bob the positions of those  rounds in which she has sent the same states. Note that Alice does not reveal any information beyond that.
    \item Bob determines $\langle\sigma_x\rangle,\langle\sigma_y\rangle,$ and $\langle\sigma_z\rangle$ for each of the states $\rho_k$ and thus determines the values of  the invariants, $I^{(r)}_1$ and $I^{(r)}_2, 1\leq r \leq k$, given in equation (\ref{eq:QSDC_qubit1}), for all the $k$ states to retrieve the message sent by Alice. These invariants carry the uncorrupted information to him. 
    \item  By employing signs of $I_1$ and $I_2$, a key consisting of binary symbols can be generated in the following manner: $I_1, I_2 >0 \to 00, ~I_1>0, I_2 <0 \to 01,~I_1<0, I_2 >0 \to 10,~I_1<0, I_2 <0 \to 11$. 
\end{enumerate}
In this way, Bob securely receives the error--free message sent by Alice even after passing through a  depolarizing noisy channel. 
 Security of this protocol against intercept-resend attack and entangle-and-measure attack is discussed below. 

In this manner, by encoding information in the invariants, error--free information can be transmitted through various noisy channels in a secure manner. Though we have shown this encoding scheme for a  QKD protocol, this template can be applied to all the information transfer protocols, e.g., quantum secure direct communication, quantum secret sharing, quantum mutual identification. In fact, the same procedure with appropriate modifications can also be opted for semi--quantum communication protocols. As another application of this encoding scheme, we have shown remote transfer of information employing qubits through a depolarising channel. In the next section, we show the emergence of standard quantum error correcting codes by putting more constraints on the choice of logical states.

\section*{Security of the proposed QKD protocol}\label{security}
 In this section, we examine the security of the protocol proposed in section ``Application" of the manuscript against intercept-resend and entangle-and-measure attacks. 

 \noindent{\bf Intercept-resend attack:}  
 Here, Eve intercepts and sends the post-measurement state to Bob. But, then, her presence will be given away with decoy states, which we recall, are sent randomly in some of the rounds. Since Eve cannot differentiate between the decoy states and the states that are employed to transfer information, she ends up measuring the decoy states. As a result of Eve's measurement, the state of system may change, which will be reflected in the sifting process when Bob reveals his measurement observables and the corresponding outcomes. Thus, the presence of Eve will be detected.  \\
 \noindent{\bf Entangle-and-Measure attack:}
 Here, Eve entangles her ancillary system with the state travelling to Bob. By performing a measurement on her system, she retrieves information on Bob's state. However if she does so, the statistics of the decoy states will change.
 
 In this way, the proposed protocol is secure against these attacks. A more general security analysis for the proposed protocols constitute a separate study and will be taken up elsewhere.

\section*{ Remote transfer of information through a depolarising channel}
Remote transfer of information requires sharing of entangled states between two parties. However, to share such states, at least one of the subsystem has to travel through channels, which are necessarily noisy. As a consequence, the state of the entire system changes. Given this situation, a good question to ask is whether information can be transferred in an error-immune manner. Again, the answer is in the affirmative if we employ the information encoding scheme proposed in the formalism of the manuscript. That is, if it is encoded in invariants for a noisy channel, it can be transferred in an error-immune manner. As a prototype, consider a maximally entangled two-qubit state with one of the qubits passing through a  depolarising channel.  

Let Alice possess a singlet state,
\begin{equation}
    \rho=\frac{1}{4}\big(\mathbb{1}-\vec{\sigma}_1\cdot\vec{\sigma}_2\big).
\end{equation}
She sends one of the two qubits to Bob. Because the channel is depolarising, the final state shared by Alice and Bob, which is corrupted,  will have the form, 
\begin{align}
    \rho' &\equiv p_0\rho+p\big(\sigma_{2x} \rho\sigma_{2x} + \sigma_{2y}\rho\sigma_{2y} +\sigma_{2z}\rho\sigma_{2z} \big)=\frac{1}{4}\big(\mathbb{1}-\alpha\vec{\sigma}_1\cdot\vec{\sigma}_2\big),~ 
\end{align}

where $\alpha=(1-4p),~ 0 \leq p_0,~ p\leq 1~ {\rm and}~ p_0+3p = 1.$
Alice wants to transmit information of her measurement to Bob without any errors. Due to the transmission of the second qubit through the noisy channel, she cannot send full information about her measurement direction. However, she can send, if not all, some partial error--free information about her measurement. This can be done by encoding information about her measurement in the invariants of a depolarising channel. This can be seen as follows:

Alice performs a measurement of an observable $\vec{\sigma}_2\cdot\hat{m}$ on the qubit in her possession. As a result of this measurement, the state of Bob will be, $\rho_B=\frac{1}{2}\big(\mathbb{1}\pm\alpha\vec{\sigma}_2\cdot\vec{m}\big)$.
Bob can retrieve information encoded in the invariants,
\begin{align}
 &I_{1}=\frac{\langle  \sigma_{2y}\rangle}{\langle  \sigma_{2x}\rangle} = \frac{m_y}{m_x},~~~~~I_{2}=\frac{\langle  \sigma_{2y}\rangle}{\langle  \sigma_{2z}\rangle} = \frac{m_y}{m_z},
\end{align}
since these are independent of $\alpha$. These invariants provide an error-free information about Alice's measurement. In this way, Alice is able to remotely transfer information about her measurement to Bob in an error-immune manner. The performance of the protocol is limited by sensitivity of the detector.


\section*{Emergence of ancilla-free quantum error correction}
\label{QECC}
\hypertarget{QECC}{}
It might appear that the method outlined in this paper is dissimilar and unrelated to the standard quantum error-correcting codes (QECC). On the contrary, the 
method is more general. In fact, by imposing additional constraints on the choice of states, standard QECC can be retrieved. In the following, we show that this is indeed the case.

As an example, consider a six-dimensional system in which dominant errors are due to the action of $X$ and $X^2$, occurring with respective probabilities of $p_1$ and $p_2$. Recall that operators $X$ and $X^2$ are $ X=\sum_{k=0}^5\ket{k+1}\bra{k},~~~~X^2=\sum_{k=0}^5\ket{k+2}\bra{k},$ with addition of integers is modulo 6.

In QECC, encoding is done in a way that each error projects the initial state onto orthogonal subspaces. This condition allows to prepare the initial state in the subspace spanned by basis states $\ket{0}$ and $\ket{3}$.   
That is to say, the state $\ket{\psi_0}$ carrying the information is given by,
\begin{equation}
    \ket{\psi_0}=\alpha\ket{0}+\beta\ket{3}.
\end{equation}
After passing through the noisy channel, the final state will have the form,
\begin{align}
    \rho &\equiv\sum_{i=0}^2p_i X^i\ket{\psi_0}\bra{\psi_0}(X^i)^\dagger= \sum_{i=0}^2p_i\ket{\psi_i}\bra{\psi_i};~~~~~ X^0\equiv \mathbb{1},
\end{align}
where the states $\ket{\psi_1}$ and $\ket{\psi_2}$ are defined as,
\begin{equation}
    \ket{\psi_1}=\alpha\ket{1}+\beta\ket{4},~~~~~~\ket{\psi_2}=\alpha\ket{2}+\beta\ket{5}.
\end{equation}
Since each of the errors $X$ and $X^2$ have projected the state onto the respective orthogonal subspaces spanned by $\{\ket{1},\ket{4}\}$ and $\{\ket{2},\ket{5}\}$, they can be discriminated unambiguously. To do so, the measurement of the following stabiliser can be performed,
\begin{equation}
    S=c_0\big(\ket{0}\bra{0}+\ket{3}\bra{3}\big)+c_1\big(\ket{1}\bra{1}+\ket{4}\bra{4}\big)+c_2\big(\ket{2}\bra{2}+\ket{5}\bra{5}\big), ~~~c_0\neq c_1 \neq c_2.
    \end{equation}
    The outcome of the measurement determines the error, whose action can be undone by applying the corresponding inverse unitary transformation. That is to say, if the outcome of measurement result is $c_0$, the state is uncorrupted whereas for the outcome $c_1(c_2)$, the error $X(X^2)$ has corrupted the state, whose effect can be mitigated by performing the inverse transformation $X^\dagger((X^2)^\dagger)$. 
Thus, by applying the appropriate unitary transformation based on the received outcomes, errors can be corrected. Note that unlike the formalism laid down in this paper, in conventional QECC codes, the basis states are fixed, which makes their implementations relatively more demanding.


\section*{Conclusion}
\hypertarget{Conclusion}{}
In summary, we have laid down a formalism to extract invariants for a number of noisy channels, which can be employed for transfer of information in an uncorrupted manner. This scheme  alleviates the need for an entangled state, whose preparation is a difficult task. It also works for mixed states, which further reduces the burden of preparation of pure states.  As applications of this encoding scheme, we have demonstrated protocols for quantum key distribution and remote transfer of information.

This work opens up a number of possibilities. For example, transfer of information using this formalism for multi-party systems constitutes an interesting study.  Multi-party states can then also be employed to transfer error-free information to multiple users under noisy communication channels. Additionally, the formalism proposed in this work can also be used to transfer error-free information in several other scenarios. For example, the information may be encoded in the invariants in such a manner that it can be retrieved only if all the parties collaborate. Besides, several communication protocols such as quantum secure direct communication, quantum secret sharing can be proposed. In conclusion, semi-quantum communication protocols can also be demonstrated for error-free information transfer through noisy channels.



\section*{Acknowledgements}

 Rajni thanks UGC for funding her research in the initial stage of the work. Sooryansh  thanks the Council for Scientific and Industrial Research (Grant no. -09/086 (1278)/2017-EMR-I) for funding his research.

\section*{Author contributions statement}
All the authors have contributed equally in all the respects at all the stages.

\section*{Data availability statement}
No data was generated in this study.

\noindent\section*{Competing interests} 
The authors declare no competing interests.

\end{document}